\documentclass[12pt]{article}
\usepackage{graphicx, amssymb, amsthm, amsbsy, amsmath}
\usepackage[normalem]{ulem}
\usepackage{multirow}
\usepackage{algorithm}
\usepackage{algpseudocode}
\usepackage{natbib}
\usepackage{epsfig}
\usepackage{subfigure}

\usepackage[margin=1in]{geometry}

\usepackage[utf8]{inputenc} 
\usepackage[T1]{fontenc}    
\usepackage{hyperref}       
\usepackage{url}            
\usepackage{booktabs}       
\usepackage{amsfonts}       
\usepackage{nicefrac}       
\usepackage{microtype}      
\usepackage{xcolor}         

\graphicspath{{fig}}

\DeclareGraphicsExtensions{.pdf,.eps.gz,.eps,.epsi.gz,.epsi,.ps,.ps.gz}
\long\def\symbolfootnote[#1]#2{\begingroup%
\def\thefootnote{\fnsymbol{footnote}}\footnote[#1]{#2}\endgroup}

\newtheorem{theorem}{Theorem}
\newtheorem{lemma}{Lemma}
\newtheorem{corollary}{Corollary}
\newtheorem{assumption}{{\bf Assumption}}

\def\A{{\bm A}}
\def\B{{\bm B}}

\def\U{{\bm U}}
\def\V{{\bm V}}
\def\W{{\bm W}}
\def\X{{\bm X}}

\def\Z{{\bm Z}}
\def\a{{\bm a}}
\def\b{{\bm b}}
\def\g{{\bm g}}
\def\v{{\bm v}}

\def\u{{\bm u}}
\def\w{{\bm w}}
\def\x{{\bm x}}
\def\y{{\bm y}}
\def\z{{\bm z}}
\def\1{{\bm 1}}
\def\0{{\bm 0}}
\def\log{\hbox{log}}
\def\eop{\hfill $\Box$}

\newcommand{\bm}{\boldsymbol}
\newcommand{\bbeta}{{\bm\beta}}
\newcommand{\bepsilon}{{\bm\epsilon}}

\newcommand{\bgamma}{{\bm\gamma}}

\newcommand{\bDelta}{{\bm\Delta}}
\newcommand{\bOmega}{{\bm\Omega}}
\newcommand{\bGamma}{{\bm\Gamma}}
\newcommand{\bSigma}{{\bm\Sigma}}

\def\mP{\bm{\mathcal{P}}}
\def\mJ{\mathcal{J}}

\def\mS{\mathcal{S}}

\def\mG{\mathcal{G}}
\def\mW{\mathcal{W}}

\newcommand{\yl}[1]{\textcolor{black}{#1}}

\title{Multi-task Learning for Gaussian Graphical Regressions with High Dimensional Covariates}
\author{{Jingfei Zhang{\small $~^{1}$} and Yi Li{\small $~^{2}$}}
\vspace{2mm}\\
\fontsize{11}{10}\selectfont\itshape
$^{1}$\,Department of Management Science, \\
\fontsize{11}{10}\selectfont\itshape
University of Miami, Coral Gables, FL 33146.\\ 
\fontsize{11}{10}\selectfont\itshape
$^{2}$\,Department of Biostatistics, \\
\fontsize{11}{10}\selectfont\itshape
University of Michigan, Ann Arbor, MI 48109.\\ 
}
\date{}

\begin{document}
\maketitle

\begin{abstract}
Gaussian graphical regression is a powerful means that regresses the precision matrix of a Gaussian graphical model on covariates, permitting the numbers of the response variables and covariates to far exceed the sample size. Model fitting is typically carried out via separate node-wise lasso regressions, ignoring the network-induced  structure among these regressions. Consequently, the error rate is high, especially when the number of nodes is large. We propose a multi-task learning estimator for fitting Gaussian graphical regression models; we design a cross-task group sparsity penalty and a within task element-wise sparsity penalty, which govern the sparsity of active covariates and their effects on the graph, respectively. For computation, we consider an efficient augmented Lagrangian algorithm, which solves subproblems with a semi-smooth Newton method. For theory, we show that the error rate of the multi-task learning based estimates has much improvement over that of the separate node-wise lasso estimates, because the cross-task  penalty  borrows information across  tasks. To address the main challenge that the tasks are entangled in a complicated correlation structure, we establish a new tail probability bound for correlated heavy-tailed (sub-exponential) variables with an arbitrary correlation structure, a useful theoretical result in its own right. Finally, the utility of our method is demonstrated through simulations as well as  an application to a gene co-expression network study with brain cancer patients. 
\end{abstract}

\section{Introduction}
Gaussian graphical models are an effective tool for inferring the dependence among variables of interest, such as the co-expression patterns among genes \citep[e.g.,][]{cai2012covariate,chen2016asymptotically} and functional connectivity between brain regions \citep[e.g.,][]{zhang2019mixed}, because  precision matrices for multivariate Gaussian variables can be interpreted as partial correlations \citep{peng2009partial}. What has been overlooked is that graphical structures may depend on external covariates, which can be high dimensional. For example,  genetic variants, clinical and environmental factors, may affect both the expression levels of individual genes and the co-expression relationships among genes \citep{wang2012snpxge2,wang2013statistical,brynedal2017large}. Genetic variants that alter co-expression relationships are referred to as co-expression quantitative trait loci (QTLs), and identifying them is of keen scientific interest \citep{wang2012snpxge2, wang2013statistical,van2018single}. Identifying co-expression QTLs among many SNPs can be formulated as a problem that relates graphical models to high dimensional external covariates, and a fundamental interest is to ascertain how the covariates modulate the individual-level graphical structures, and recover both the population- and subject-level graphs. 

Though much progress has been made  for developing graphical models, less has been done for covariate-dependent graphical models.
Several works \citep[e.g.,][]{yin2011sparse,li2012sparse,cai2012covariate,chen2016asymptotically} considered covariate-dependent Gaussian graphical models, wherein the mean of the nodes depends on covariates, while the network structure is the same across all of the subjects.
\cite{guo2011joint} and \cite{danaher2014joint} estimated  stratified graphical models by preserving the common structure among them. 
\cite{liu2010graph} partitioned the covariate space into several subspaces and fitted separate Gaussian graphical models for each subspace. 
\cite{kolar2010sparse}  nonparametrically estimated the dependence of a covariance matrix on one continuous covariate. 
\cite{cheng2014sparse} fitted a conditional Ising model for binary data. 
\cite{ni2019bayesian} proposed a DAG model that allows the graph structure to vary with a small number of covariates, and assumed a  hierarchical ordering of the nodes. None of the cited works, however, can handle high dimensional covariates.
Notably, \cite{zhang2022high} proposed a Gaussian graphical regression framework that regresses the precision matrix of a graphical model on high dimensional covariates and estimates the parameters using penalized node-wise regressions. However, \cite{zhang2022high} utilized separate node-wise regressions, ignoring the network-induced common structure among these regressions. Consequently, the obtained error rate is high, especially with a large number of nodes.

This work addresses these issues by making several methodological and theoretical contributions.
\begin{itemize}
\item We propose to adapt multi-task learning to fit Gaussian graphical regression models. Specifically, 
we design a cross-task group sparsity penalty  which enables us to borrow information across different tasks. As the number of covariates, $q$, can be much larger than $n$, it is reasonable to assume the active covariates (i.e., those with nonzero effects on the graph) are sparse, which can also be accommodated by the specified cross-task group sparsity penalty; see Figure \ref{fig:illu}.
We  combine it with an element-wise sparsity to address the possible sparse effects of active covariates and further reduce the model complexity. 

\item To optimize the cross-task objective function with the combined sparsity penalty, we adapt an efficient augmented Lagrangian algorithm, which solves subproblems with a semi-smooth Newton method. 

\item To address the main challenge that the regression tasks are entangled in a general correlation structure and that the combined sparsity penalty is not decomposable, we establish a new and sharp tail probability bound for correlated heavy-tailed (sub-exponential) variables with an arbitrary correlation structure, which is meritorious even on its own.
\item Finally, we prove that, compared to \citet{zhang2022high}, the error rate of the simultaneously estimated precision parameters  improves by a factor of $p$, the number of response variables.  The improvement is remarkable for a large $p$, as  further corroborated in simulation studies.  Code is available on GitHub, and detailed  proofs are given in the Supplementary Material.
\end{itemize}

\section{Multi-task Learning for Gaussian Graphical Regressions}
Denote by $\X=(X_1,\ldots,X_p)^\top$ the vector of response variables (e.g., expression levels of $p$ genes) and $\U=(U_1,\ldots,U_q)^\top$ the vector of covariates (e.g., age, sex and genetic variants), and assume that 
$$
\X\,\vert\, \U=\u  \, \sim \,  \mathcal{N}_p(\bGamma\u,\bSigma(\u)),
$$
where $\bSigma(\u)$ is the conditional covariance, and $\bOmega(\u)=\bSigma^{-1}(\u)$ is the conditional precision matrix linked to $\u$ via
\begin{equation}\label{eqn:igmm2}
-\bOmega(\u)_{jk}=
\begin{cases}
-\sigma^{jj}     & \quad j=k,\\
\beta'_{jk0}+\sum_{h=1}^q\beta'_{jkh}u_h & \quad j\neq k,\\
 \end{cases}
\end{equation}
where $\beta'_{jkh}=\beta'_{kjh}$, $h\in\{0\}\cup[q]$, $j,k\in[p]$. 
We assume $\bOmega(\u)_{jj}=\sigma^{jj}$ {to be free of $\u$}, and this is discussed after \eqref{reg}. Writing $\Z=\X-\bGamma\u=(Z_1, \ldots, Z_p)^\top$, some straightforward algebra shows that (\ref{eqn:igmm2}) relates $\bOmega(\u)$ to the following regression, termed \textit{Gaussian graphical regression} \citep{zhang2022high}:
\begin{equation}\label{reg}
Z_j={\sum}_{k\neq j}^p\beta_{jk0}Z_k+{\sum}_{k\neq j}^p{\sum}_{h=1}^q\underbrace{\beta_{jkh}\times u_hZ_k}_{\text{{interaction term}}} +\epsilon_j,
\end{equation}
where $\beta_{jkh}=\beta'_{jkh}/\sigma^{jj}$, $\epsilon_j$ is independent of $\Z_{-j}=\{Z_k:k\neq j\in[p]\}$ and $\text{Var}(\epsilon_j)=1/\sigma^{jj}$, for all $j,k$ and $h$.
It is easy to see that, when $\beta_{jkh}=0$ for all $j,k,h\neq 0$, \eqref{reg} reduces to the usual Gaussian graphical model with $\Z\,\vert\, \U=\u  \, \sim \,  \mathcal{N}_p(\0,\bSigma)$.

\begin{figure}[!t]
	\centering
	\includegraphics[trim=0 0 0 0, scale=0.8]{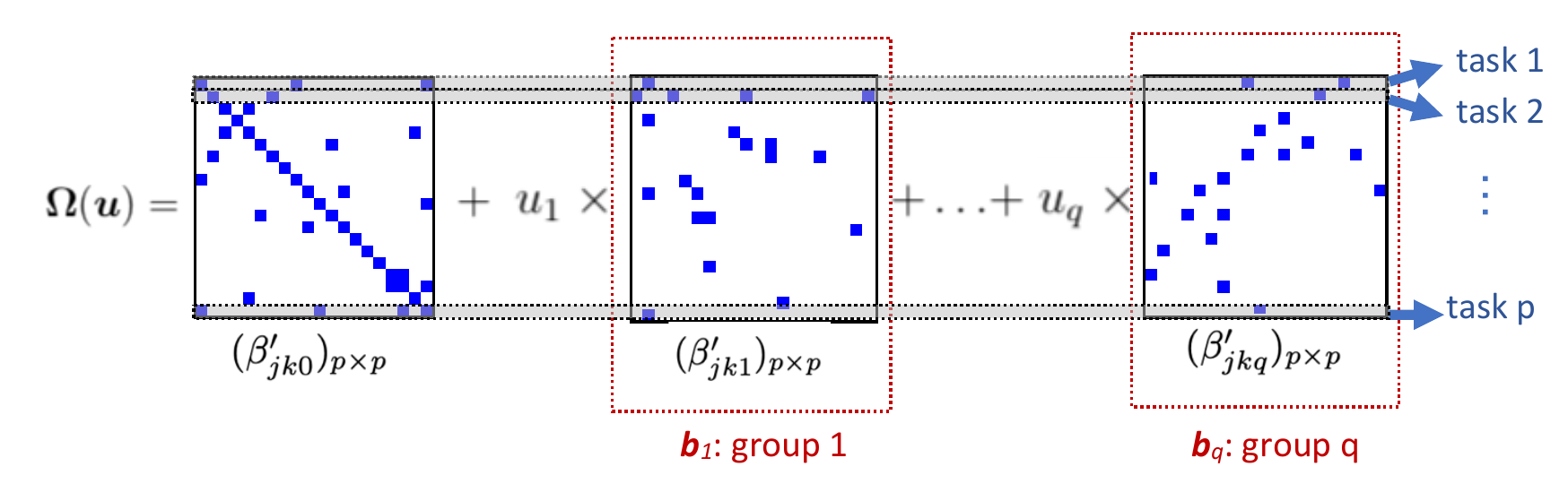}
	\caption{An illustration of multi-task learning in Gaussian graphical regression.}
	\label{fig:illu}
\end{figure}

Model \eqref{reg} provides a regression framework for estimating the precision parameters in  \eqref{eqn:igmm2}, by including the interactions between $\X_{-j}$ and $\u$. Correspondingly, the partial correlation between $X_j$ and $X_k$, conditional on all other $X$ variables, {is modeled as} a function of $\u$; see an illustration in Figure \ref{fig:illu}.
The diagonal elements of $\bOmega(\u)$ (i.e., $\sigma^{jj}$'s) are connected to the residual variances in \eqref{reg}, that is, $\text{Var}(\epsilon_j)=1/\sigma^{jj}$. From this perspective, assuming $\sigma^{jj}$ to be free of $\u$ may be viewed as assuming the residual variance of $Z_j$, after removing the effects of $\u$, $\Z_{-j}$ and their interactions, not to depend on $\u$, which is plausible in the context of regression.

Let $\bbeta_j=(\beta_{j10},\ldots, \beta_{j,j-1,0}, \beta_{j,j+1,0}, \ldots, \beta_{jp0},\ldots,\beta_{j1q},\ldots, \beta_{j,j-1,q}, \beta_{j,j+1,q}, \ldots,\beta_{jpq})^\top\in\mathbb{R}^{(p-1)(q+1)}$ be the vector of coefficients in \eqref{reg} and denote $\bbeta=(\bbeta_1,\ldots,\bbeta_p)$. To expose the key ideas, we assume a known $\bGamma$ in the ensuing development, and focus on the estimation of $\bbeta$;
extensions with unknown $\bGamma$ are possible, but with more involved notation.
With $n$ independent data $\mathcal{D}=\{(\u^{(i)},\x^{(i)}), i\in[n]\}$, let $\z^{(i)}=\x^{(i)}-\bGamma\u^{(i)}$ and $\w^{(i)}_{-j}=\z_{-j}^{(i)}\otimes \u^{(i)}$, where $\otimes$ denotes the Kronecker product. To estimate $\bbeta$, we consider
\begin{equation}\label{eqn:obj}
\arg{\min}_{\bbeta
}\frac{1}{2n}\sum_{j=1}^p\sum_{i=1}^n(z^{(i)}_j-\w^{(i)}_{-j}{}^\top\bbeta_j)^2+\lambda_1\sum_{h=1}^q\Vert\b_h\Vert_2+\lambda_2\Vert\bbeta\Vert_1,
\end{equation}
where $\b_h=((\bbeta_1)_{(h)},\ldots,(\bbeta_p)_{(h)})$ with $(\bbeta_j)_{(h)}=(\beta_{j1h},\ldots, \beta_{j,j-1,h},  \beta_{j,j+1,h}, \ldots, \beta_{jph})$ and $\lambda_1,\lambda_2\ge0$ are tuning parameters. 
The convex regularizing terms, $\sum_{h=1}^q\Vert\b_h\Vert_2$ and $\Vert\bbeta\Vert_1$, encourage group- and element-wise sparsity, respectively, though the group sparse penalty is not applied to $(\bbeta_j)_{(0)}$ (i.e., intercept terms). The combined sparsity
penalty in \eqref{eqn:obj} is termed the {\it sparse group lasso} penalty \citep{simon2013sparse}.
Distinguishing from the node-wise estimation \citep{zhang2022high},
\eqref{eqn:obj}  considers $p$ graphical regressions simultaneously, with the group lasso penalty regulating $\b_h$ (the effects from the $h$th covariate) cross tasks; see  Figure \ref{fig:illu}. Thus, \eqref{eqn:obj} is termed multi-task learning for a Gaussian graphical regression model.

\section{An Augmented Lagrangian Algorithm with a Semi-smooth Newton Method }
\label{sec:est}
By specifying a convex objective function,  \eqref{eqn:obj} facilitates use of
 existing algorithms for solving sparse group lasso problems,
including proximal gradient descent \citep{chen2012smoothing}, block coordinate descent \citep{simon2013sparse, wang2019two} and alternating direction method of multipliers (ADMM) \citep{boyd2011distributed}. 
We  propose to use an efficient augmented Lagrangian algorithm which uses a semi-smooth Newton method to solve subproblems. {We opt for the semi-smooth Newton method because it exploits the second order sparsity structure in our setting \citep{zhang2020efficient} and }
enjoys a  linear rate of convergence. 

More  specifically, write $\mW= \begin{pmatrix}
 \W_{-1} & \cdots & \0_{n\times (p-1)(q+1)} \\
 \vdots  & \ddots & \vdots  \\
 \0_{n\times (p-1)(q+1)} & \cdots & \W_{-p}
 \end{pmatrix}$, where $\W_{-j}=[\w^{(1)}_{-j};\ldots;\w^{(n)}_{-j}]$, and $\a=(z^{(1)}_1,\ldots,z^{(n)}_1,z^{(2)}_1,\ldots,z^{(n)}_p)\in\mathbb{R}^{np}$.
Correspondingly, \eqref{eqn:obj} can be written as 
$$
\arg{\min}_{\bbeta}\frac{1}{2n}\|\mW\bbeta-\a\|_2^2+\lambda_1\sum_{h=1}^q\Vert\b_h\Vert_2+\lambda_2\Vert\bbeta\Vert_1,
$$ 
and its dual problem can be written as:
\begin{eqnarray*}
&&\min_{\y,\g}\,\, \langle\a,\y\rangle +\frac{1}{2}\|\y\|^2+h^*(\g)\\
&&\text{s.t.}\,\, \mW^\top\y+\g=0,
\end{eqnarray*}
where $h^*(\cdot)$ is the conjugate function of $h(\bbeta):=\lambda_1\sum_{h=1}^q\Vert\b_h\Vert_2+\lambda_2\Vert\bbeta\Vert_1$, {i.e., $h^*(\bbeta)=\sup_{\bm{e}}\{\langle\bbeta,\bm{e}\rangle-h(\bm{e})\}$}.
Define the proximal mapping of $h(\cdot)$  to be 
$$
\text{Prox}_{h}(\u)=\arg\min_\x\left\{h(\x)+\frac{1}{2}\|\x-\u\|^2\right\}.
$$
Similar to ADMM, 
the augmented Lagrangian of the dual problem is given by 
$$
\langle\a,\y\rangle +\frac{1}{2}\|\y\|^2+h^*(\g)+\frac{\tau}{2}\|\mW^\top\y+\g-\sigma^{-1}\bbeta\|^2-\frac{1}{2\tau}\|\bbeta\|^2,
$$
where $\tau>0$, and the subproblem relates to iteratively minimizing  
$$
\psi_k(\y)=\langle\a,\y\rangle+\frac{1}{2}\|\y\|_2^2+h^*(\text{Prox}_{h^*/\tau_k}(\tau_k^{-1}\bbeta^k-\mW^\top\y)+\frac{\tau_k}{2}\|\text{Prox}_{h}(\tau_k^{-1}\bbeta^k-\mW^\top\y)\|_2^2-\frac{1}{2\tau_k}\|\bbeta^k\|^2,
$$
where $\bbeta^k$ is the updated estimate and $\tau_k>0$ is given at the $k$th iteration; see the detailed steps below.

\begin{algorithm}
\caption{An augmented Lagrangian method for solving \eqref{eqn:obj}}
\begin{algorithmic}
\State Let $\tau_0 > 0$ be given, and  choose $(\y_0,\g_0, \bbeta_0)$. Iterate the following steps for $k = 0, 1, \ldots,$ until convergence.
\State Step 1: compute $\y^{k+1}= \min_\y\psi_k(\y)$ and 
$\g^{k+1}=\text{Prox}_{p^*/\tau_k}(\tau_k^{-1}\bbeta^k-\mW^\top\y^{k+1})$;
\State Step 2: compute $\bbeta^{k+1}=\bbeta^k-\tau_k(\mW^\top\y^{k+1}+\g^{k+1})$;
\State Step 3: update $\tau_{k+1}$ so that $\tau_{k+1} > \tau_{k}$.
\end{algorithmic}
\label{algo}
\end{algorithm}
Step 1 solves the subproblem $\min_\y\psi_k(\y)$ with a semi-smooth Newton method, utilizing a generalized Jacobian of $\text{Prox}_{h}(\cdot)$ given in \cite{zhang2020efficient}.
In general, the algorithm outperforms the  first-ordered ADMM algorithm in  computational efficiency and estimation accuracy, because it accounts for the second order sparsity structure when solving  $\min_\y\psi_k(\y)$ \citep{li2017semi,zhang2020efficient}.


\textbf{Positive-definiteness.} 
A natural sufficient condition to ensure  positive definiteness of $\bOmega(\u)$ is  diagonal dominance, leading to $\max(1,\Vert\u\Vert_{\infty})\Vert\bbeta_j\Vert_1 < 1$.
With Assumption \ref{ass1} stipulating $|u^{(i)}_h|\le M$, positive definiteness holds when $\Vert\bbeta_j\Vert_1 < 1/\max(1, M)$, $j\in[p]$. 
Assuming $u_h\in[-1,1]$ (if not, rescale first), this sufficient condition can be simplified to $\Vert\bbeta_j\Vert_1 < 1$ for all $j$ (note that $\bbeta_j$ is sparse), suggesting that, to satisfy diagonal dominance, the ``effect sizes"  of $\u$  (i.e., $\|\bbeta_j\|_1$) on partial correlations cannot be too large. If the true covariance/precision matrix is positive definite,  Theorem \ref{thm1} implies that the estimated precision matrix is asymptotically positive definite. For finite sample cases, a post-hoc rescaling procedure can be utilized for securing  positive definiteness.

\textbf{Tuning.} The $\lambda_1$ and $\lambda_2$ in \eqref{eqn:obj} can be jointly selected via $L$-fold (e.g., $L=5$) cross validation. To facilitate parallelization of the algorithm, we consider a hierarchical selection procedure.
Rewrite $\lambda_1=(1-\alpha)\lambda_0$ and $\lambda_2=\alpha\lambda_0$, where $\alpha $ reflects the weight of the lasso penalty relative to the group lasso penalty 
(e.g., $\alpha=0$ and $1$ correspond to group lasso and lasso, respectively)
and $\lambda_0$ reflects the total amount of regularization. We assess a set of values for $\alpha\in[0,1]$; for each $\alpha$, {a grid of} $\lambda_0$ values are considered in cross validation. The search can be parallelized for different $\alpha$'s.


\section{Theory: Concentration Inequality and Error Rate for Multi-task Learning}
\label{sec:theory}
We derive the non-asymptotic $\ell_2$ error rate of the sparse group lasso estimator from \eqref{eqn:obj}. The main challenge is that the $p$ tasks in \eqref{eqn:obj} are correlated,  
and this differs from the usual multi-task learning with group sparsity \citep{lounici2011oracle}.
To tackle this challenge, we have made a key advance in Theorem \ref{thm0} that gives a new tail bound for the sum of correlated heavy-tailed (sub-exponential) variables with an arbitrary correlation structure. 

Our theoretical investigation faces other challenges.
First, because the design matrix in \eqref{eqn:obj} includes high-dimensional interactions between $\z^{(i)}$ and $\u^{(i)}$, and the variance of $\z^{(i)}$ is a function of $\u^{(i)}$, characterizing the joint distribution of each row in $\W_{-j}$ is difficult and requires a delicate treatment.  
Second, as the combined penalty term $\lambda_1\sum_{h=1}^q\Vert\b_h\Vert_2+\lambda_2\Vert\bbeta\Vert_1$ is not decomposable, the classic techniques for decomposable regularizers and null space properties \citep{negahban2012unified} are not applicable. 
By utilizing a novel concentration inequality for the sum of correlated variables in \eqref{eqn:obj},
we derive two interrelated bounds for the stochastic term, whose combination
yields a sharp upper bound of the stochastic term. We, therefore, show that our proposed estimator possesses an improved $\ell_2$ error bounds compared to the lasso and the group lasso when the true coefficients are simultaneously sparse, and, more importantly,  the error rate of the multi-task estimates  improves by a factor of $p$, compared to the separate  node-wise estimates obtained by \citet{zhang2022high}.

\subsection{A new concentration inequality for the sum of correlated variables}
The concentration results for
correlated random variables are often derived under  specific correlation structures, such as  weak dependence \citep{merlevede2011bernstein} and asymptotic independence \citep{ko2008sums}. These structures are unlikely to be applicable to our setting because the error terms in \eqref{eqn:obj}, $\epsilon_1,\ldots,\epsilon_p$ (from the $p$ tasks), depend on, for example, the expressions of $p$ genes, and are correlated via a complicated co-expression network. To bound $\sum_j\epsilon_j$, we employ a novel idea that partitions the index set of $p$ response variables into mutually exclusive subsets such that the variables within each subset are independent. This can be done by exploring the
topology of a graph and solving a vertex coloring problem \citep{lewis2015guide}; see the proof of Theorem \ref{thm0} in the Supplementary Material. With that, we present a  concentration inequality result under a  general correlation structure. 

\begin{theorem}\label{thm0}
Consider $N$ correlated mean zero sub-exponential random variables $Y_j$, $j\in[N]$ and an induced network $G(V,E)$ with a node set $V=\{1,\ldots,N\}$ and an edge set $E= \{ (j, k):\text{Cov}(Y_j,Y_k)\neq0 \}$. Denote the maximum node degree of $G(V,E)$ by $d_{\max}$ and let $c_G=\min\left(d_{\max}+1,\frac{1+\sqrt{8|E|+1}}{2}\right)$. For any $t\ge 0$ and a constant $c>0$, it holds that
$$
\mathbb{P}\left(\sum_{j=1}^NY_j\ge t\right)\le c_G\exp\left[-c\min\left\{\frac{t^2}{c_G^2\sum_{j=1}^N\Vert Y_i\Vert_{\psi_1}^2},\frac{t}{c_G\max_j\Vert Y_i\Vert_{\psi_1}}\right\}\right],
$$
\end{theorem}
\yl{where $\|\cdot\|_{\psi_1}$ is the sub-exponential norm  defined in the Supplementary Material.}
The results are inclusive. For example, when $Y_1,\ldots,Y_N$ are mutually independent, we have $c_G=1$, and the inequality reduces to the usual form for independent sub-exponential random variables; if the variables are correlated with $c_G=O(1)$, we obtain a tail probability in the same order as when the variables are independent. The theorem leads to a corollary that gives a sharp bound on the sum of correlated chi-squared random variables, which is critical for our proof.
\begin{corollary}\label{lemma8}
Consider $N$ correlated $\chi^2$ variables $Y_j\sim\chi^2_{d_j}$, $j\in[N]$ and an induced network $G(V,E)$ defined as in Theorem \ref{thm0}. Denote the maximum node degree of $G(V,E)$ by $d_{\max}$ and let $c_G=\min\left(d_{\max}+1,\frac{1+\sqrt{8|E|+1}}{2}\right)$. For any $t\ge 0$, it holds that
$$
\mathbb{P}\left(\sum_{j=1}^NY_j-\sum_{j=1}^Nd_j\ge t\right)\le c_G\exp\left[-\frac{\left\{t-(c_G-1)\sum_{j=1}^Nd_j\right\}^2}{4c_G(t+\sum_{j=1}^Nd_j)}\right].
$$
\end{corollary}

\subsection{Error rate analysis}

Let $\mS$ be the element-wise support set of $\bbeta$ and $\mG$ be the group-wise support set of $\bbeta$, and denote by $s_e=|\mS|$ and $s_{g}=|\mG|$, i.e., $s_e$ and $s_g$ are the numbers of nonzero entries and  nonzero groups, respectively. Without loss of generality, we assume $\sigma^{jj}=1$. We state the needed regularity conditions. 
\begin{assumption}
\label{ass1}
Suppose $\u^{(i)}$ are i.i.d. mean zero random vectors with a covariance matrix satisfying $\lambda_{\min}(\text{Cov}(\u^{(i)}))\ge 1/\phi_0$ for some constant $\phi_0>0$. Moreover, there exists a constant $M>0$ such that $|u^{(i)}_h|\le M$ for all $i$ and $h$.
\end{assumption}

\begin{assumption}
\label{ass2}
Suppose $\phi_1\le\lambda_{min}(\text{Cov}(\z^{(i)}))\le\lambda_{max}(\text{Cov}(\z^{(i)}))\le\phi_2$ for some constants $\phi_1,\phi_2>0$.
\end{assumption}
\begin{assumption}
\label{ass3}
The dimensions $p,q$ and sparsity $s_e$ satisfy $\log\,p+\log\,q=\mathcal{O}(n^\delta)$ and $s_e=o(n^\delta)$ for $\delta\in[0,1/6]$. 
The maximum column $\ell_0$ norm of $\bOmega(\u)$ is bounded above by a positive constant $d_0$.
\end{assumption}

Assumption \ref{ass1} stipulates that the covariates are element-wise bounded, which is needed in characterizing the joint distribution of each row in $\mW$. This condition is not restrictive as genetic variants are often coded to be $\{0,1\}$ or $\{0,1,2\}$ \citep{chen2016asymptotically}. 
Assumptions \ref{ass1} and \ref{ass2} impose bounded eigenvalues on $\text{Cov}(\u^{(i)})$ and $\text{Cov}(\z^{(i)})$ as commonly done in the high-dimensional regression literature \citep{chen2016asymptotically,hao2018model,cai2019sparse}. 
Assumption \ref{ass3} is a sparsity condition, allowing both $\log\,p$ and $\log\,q$ to grow at a polynomial order of $n$. Moreover, the number of nonzero entries $s_e$ can  grow with $n$. This condition and $\delta\in[0,1/6]$ are useful when establishing a {restricted eigenvalue} condition \citep{bickel2009simultaneous} for $\mW^\top\mW/n$ and when bounding the stochastic term $\langle\bepsilon,\mW\Delta\rangle$.
The condition on the column $\ell_0$ norm of $\bOmega(\u)$ implies that, for example, each gene is only connected to a finite number of genes.


\begin{theorem}\label{thm1}
Let $s_{\lambda}$ denote the number of nonzero entries in a candidate model such that $s_e<s_{\lambda}\le n$.
Suppose Assumptions \ref{ass1}-\ref{ass3} hold, $s_{\lambda}(\log\,p+\log\,q)=\mathcal{O}(\sqrt{n})$ and $n\ge A_1\{s_g\log(eq/s_g)+s_{e}\log(ep)\}$ for some constant $A_1>0$.
Then the sparse group lasso estimator $\hat\bbeta$ in \eqref{eqn:obj} with 
\begin{equation}\label{eqn:lambda}
\lambda_1=C\sqrt{\log(eq/s_g)/n+2s_{e}\log(ep)/(ns_g)},\quad \lambda_2=\sqrt{s_g/s_{e}}\lambda_1
\end{equation}
satisfies, with probability at least $1-C_1\exp[-C_2\{s_g\log(eq/s_g)+s_{e}\log(ep)\}]$, 
\begin{equation}\label{eqn:bound1}
\|\hat\bbeta-\bbeta\|_2^2\precsim \frac{1}{n}\{s_g\log(eq/s_g)+s_{e}\log(ep)\},
\end{equation}
where $C$, $C_1$, and $C_2$ are positive constants.
\end{theorem}

The separate node-wise regressions considered in \citet{zhang2022high} yield an error rate of $\frac{1}{n}\{ps_g\log(eq/s_g)+s_{e}\log(ep)\}$, which is slower than that in \eqref{eqn:bound1} by a factor $p$ if $s_{e}\log p=O(s_g\log q)$; 
as  $p$ often far exceeds $n$, the improvement with  multi-task learning is  considerable.

Theorem \ref{thm1} also shows that our proposed estimator enjoys an improved $\ell_2$ error bound over {estimators with only a lasso or a group lasso penalty on $\bbeta$.}
Specifically, given that the dimension of $\bbeta$ is $p(p-1)(q+1)$ and $s_{g}\le s_e$, applying the regular lasso regularizer $\lambda\Vert\bbeta\Vert_1$ alone would yield an error bound of $(s_e/n)\log(pq)$ \citep{negahban2012unified}, which is slower than that in \eqref{eqn:bound1} when $\log p/\log q=o(1)$ and $s_{g}/s_e=o(1)$, {corresponding to group sparsity}. 
Moreover, when $p>n+1$, estimating with the group lasso regularizer $\lambda_1\sum_h\Vert\b_h\Vert_2$ alone, which excludes $(\bbeta_j)_{(0)}$, is not feasible, because the dimension of the latter (i.e., $p(p-1)$) exceeds $n$. If we utilize a group lasso regularizer $\lambda_{g}\Vert\bbeta_{j}\Vert_{1,2}$ that includes $\b_0$, the estimator would have an $\ell_2$ error bound of $(s_{g}/n)\log\,q+(s_{g}/n)p(p-1)$ \citep{lounici2011oracle}, which is slower than that in \eqref{eqn:bound1} when $\log q/\{p(p-1)\}=o(1)$ and $s_j/ s_{j,g}=o(p(p-1)/\log p)$, corresponding to within-group sparsity. 
In Theorem \ref{thm1}, the condition $s_{\lambda}(\log\,p+\log\,q)=\mathcal{O}(\sqrt{n})$ upper bounds the size of candidate models, which  helps to bound the stochastic errors from $p$ tasks. 

\section{Numerical Experiments}
\label{sec:num}
We compare the finite sample performance of our proposed method,  defined in \eqref{eqn:obj} (referred to as ``\texttt{MtRegGMM}"), 
 with those of three competing solutions, namely,
a benchmarking i.i.d Gaussian graphical model estimated by the neighborhood selection method  \citep{meinshausen2006high} (``\texttt{IID}"), a lasso estimator 
$\arg{\min}_{\bbeta
}\frac{1}{2n}\sum_{j=1}^p\sum_{i=1}^n(z^{(i)}_j-\W^{(i)}_{-j}{}^\top\bbeta_j)^2+\lambda\Vert\bbeta\Vert_1$ (\texttt{Joint$_{\text{lasso}}$}), and  the separate regressions considered in \citet{zhang2022high}  (``\texttt{RegGMM}").

We simulate $n$ samples $\{(\u^{(i)},\x^{(i)}), i\in[n]\}$ from  \eqref{eqn:igmm2}, with $\x^{(i)}\in\mathbb{R}^p$ (e.g., genes) and external covariate $\u^{(i)}\in\mathbb{R}^q$ (e.g., SNPs). The $\u^{(i)}_j$'s are generated i.i.d. from Bernoulli$(0.5)$. We randomly select 3 covariates to have nonzero effects, and in the graphs for these covariates and the population-level graph, five edges are randomly selected to be nonzero.
We set $\sigma^{jj}=1$ for $j\in[p]$. The initial nonzero coefficients $\beta_{jkh}$ are set to 0.3. 
For each $j$, we  rescale $\{\beta_{jkh}\}_{k\neq j\in[p], h\in\{0\}\cup[q]}$ by dividing each entry by $\sum_{k\neq j\in[p], h\in\{0\}\cup[q]}|\beta_{jkh}|$. 
{After rescaling, for each $j,k$ and $h$, we use the average of $\beta_{jkh}$ and $\beta_{kjh}$ to fill the entries at $jkh$ and  $kjh$.} 
This process results in symmetry with diagonal dominance and, thus,  ensures  positive definiteness of the precision matrices. 
For each simulation configuration, we generate 50 independent data sets; 
given $\u^{(i)}$, we  determine $\bOmega(\u^{(i)})$ and $\bSigma(\u^{(i)})$, and generate the $i$th sample $\x^{(i)}$ from $\mathcal{N}(\bGamma\u^{(i)},\bSigma(\u^{(i)}))$, $i\in[n]$.
For a fair comparison, tuning parameters in all of the methods are selected via 5-fold cross validation.

\begin{table}[!t]
\setlength{\tabcolsep}{3pt}
\centering
\caption{Estimation accuracy of $\bbeta_j$ and $\bOmega_j$ in simulations with varying sample size network size $p$ and covariate dimension $q$. }\label{tab1}
{\renewcommand{\arraystretch}{1.15}
\begin{tabular}{|c|c|c|cccc|} \hline
$n$ &  $p$, $q$  &  Method   & TPR$_{\bbeta}$ & FPR$_{\bbeta}$ & Error of $\bbeta$ & Error of $\bOmega$ \\\hline
\multirow{12}{*}{100} 
& \multirow{4}{*}{\begin{tabular}[c]{@{}l@{}} $p=20$\\ $q=100$\end{tabular}} 
&   \texttt{MtRegGMM}   &  {\bf 0.985} (0.013)       & {\bf 0.0001} (0.0000)   & {\bf 0.534} (0.032)      & {\bf 0.405} (0.046)     \\
&& \texttt{Joint$_{\text{lasso}}$}         & 0.942 (0.015)  & 0.0003 (0.0000)   &  0.762 (0.033)            & 0.712 (0.067) \\
&& \texttt{RegGMM}            & 0.922 (0.020)  &  0.0013 (0.0001)   & 1.146 (0.045)  &  1.984 (0.153) \\
&& \texttt{IID}            & - & -   & -   & 1.764 (0.034) \\\cline{2-7}

& \multirow{4}{*}{\begin{tabular}[c]{@{}l@{}}$p=20$\\ $q=200$\end{tabular}} 
&  \texttt{MtRegGMM}    & {\bf 0.983} (0.021)   & {\bf 0.0001} (0.0000)   & {\bf 0.549} (0.037)       &  {\bf 0.436} (0.060)        \\
&&   \texttt{Joint$_{\text{lasso}}$}    & 0.918 (0.030)   &   0.0002 (0.0000)   & 0.848 (0.040)       &  0.912 (0.088)          \\
&& \texttt{RegGMM}          & 0.923 (0.016)         & 0.0009 (0.0001)   & 1.207 (0.036)     & 2.223 (0.199) \\
&& \texttt{IID}            & - & -   & -   & 1.777 (0.026) \\\cline{2-7}

& \multirow{4}{*}{\begin{tabular}[c]{@{}l@{}} $p=100$\\ $q=200$\end{tabular}} 
&  \texttt{MtRegGMM}            & {\bf 0.964} (0.013)         & {\bf 0.0000} (0.0000)   & {\bf 0.612 (0.034)}       & {\bf 0.489} (0.055) \\
&&   \texttt{Joint$_{\text{lasso}}$}   & 0.896 (0.020)   & {\bf 0.0000} (0.0000)   & 0.846 (0.038)       &  0.786 (0.069)           \\
&& \texttt{RegGMM}     &  0.918 (0.020)        & 0.0003 (0.0000)         & 2.620 (0.060)            & 5.845 (0.329) \\
&& \texttt{IID}            & - & -   & -   & 5.119 (0.890)  \\\hline
\end{tabular}}
\end{table}
\begin{figure}[!t]
\centering
\includegraphics[trim=0 5mm 0 0mm, scale=0.425]{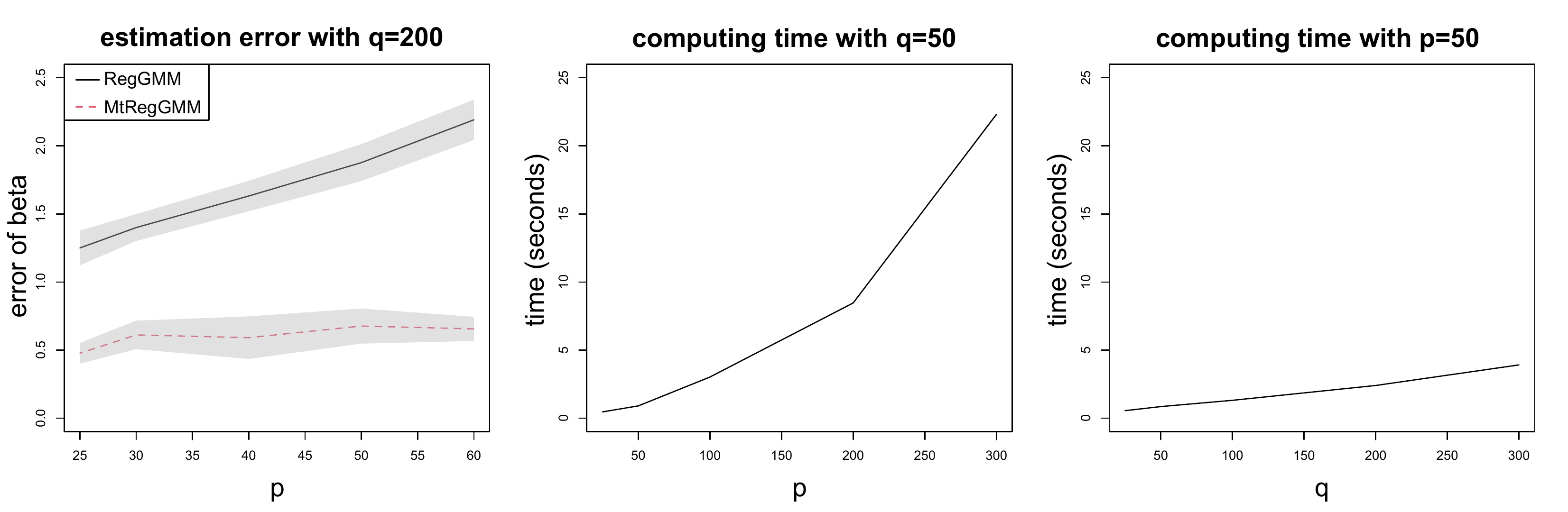}
\caption{Estimation errors and computing time.}
\label{fig1}
\end{figure}

To evaluate the estimation accuracy, we report the estimation errors  $\|{\hat\bbeta-\bbeta} \|_2$, {where $\hat\bbeta$'s, with a slight overuse of notation, denote the estimates of $\bbeta$'s obtained by various methods}. For the selection accuracy, we report the true positive rate (TPR) and false positive rate (FPR). Also reported is the average estimation error of the precision matrix defined to be $\sum_{i=1}^n\Vert\hat{\bOmega}_i-\bOmega_i\Vert_{F,\text{off}}^2/n$, where $\bOmega_i$
is the true subject-specific precision matrix
and $\hat\bOmega_i$ is estimated by a given method. 

The proposed \texttt{MtRegGMM} outperforms the competing methods in estimation and selection accuracy for various $p$ and $q$. The estimation errors of \texttt{MtRegGMM} increases with $p$ and $q$, confirming the  results of Theorem \ref{thm1}.
In the left plot of Figure \ref{fig1}, we compare the estimation errors from \texttt{RegGMM} and \texttt{MtRegGMM} as $p$ increases. The error from \texttt{RegGMM} increases roughly linearly with $p$ while that from \texttt{MtRegGMM} remains relatively stable as $p$ increases, again confirming the results of Theorem \ref{thm1}.

Finally, we assess the computation cost including tuning. The middle and right panels of Figure \ref{fig1} show the computation time of \texttt{MtRegGMM} for a given $(\lambda_1,\lambda_2)$. The simulations were run on an iMac with a 3.6 GHz Intel Core i9 processor. Because the number of parameters is $\mathcal{O}(p^2q)$, the total computing cost is expected to be in the order of $p^2q$,
as seen in Figure \ref{fig1}.

\section{Co-expression QTL Analysis} 

We analyze the REMBRANDT trial (GSE108476) with a subcohort of $n=178$ glioblastoma multiforme (GBM) patients, who had undergone microarray and single nucleotide polymorphism (SNP) chip profiling, with both gene expression and SNP data available for analysis. The raw data were pre-processed and normalized using standard pipelines \citep{gusev2018rembrandt}. 

The response variables are the expression levels of $p=73$ genes  belonging to the human glioma pathway in the Kyoto Encyclopedia of Genes and Genomes (KEGG) database \citep{kanehisa2000kegg}. The covariates  include local SNPs (i.e., SNPs that fall within 2kb upstream and 0.5kb downstream of the gene) residing near these 73 genes, resulting in a total of 118 SNPs. SNPs are coded with ``0" indicating homozygous in the major allele and ``1"  otherwise.  Age and gender are also included in analysis. Consequently, there are $q=120$ covariates, bringing a total of $73\times36\times121 = 317,988$  parameters (including intercepts). We construct the population-level gene co-expression network, and examine if and how age, gender and SNPs modulate the network.

\begin{figure}[!t]
\centering
\includegraphics[scale=0.4]{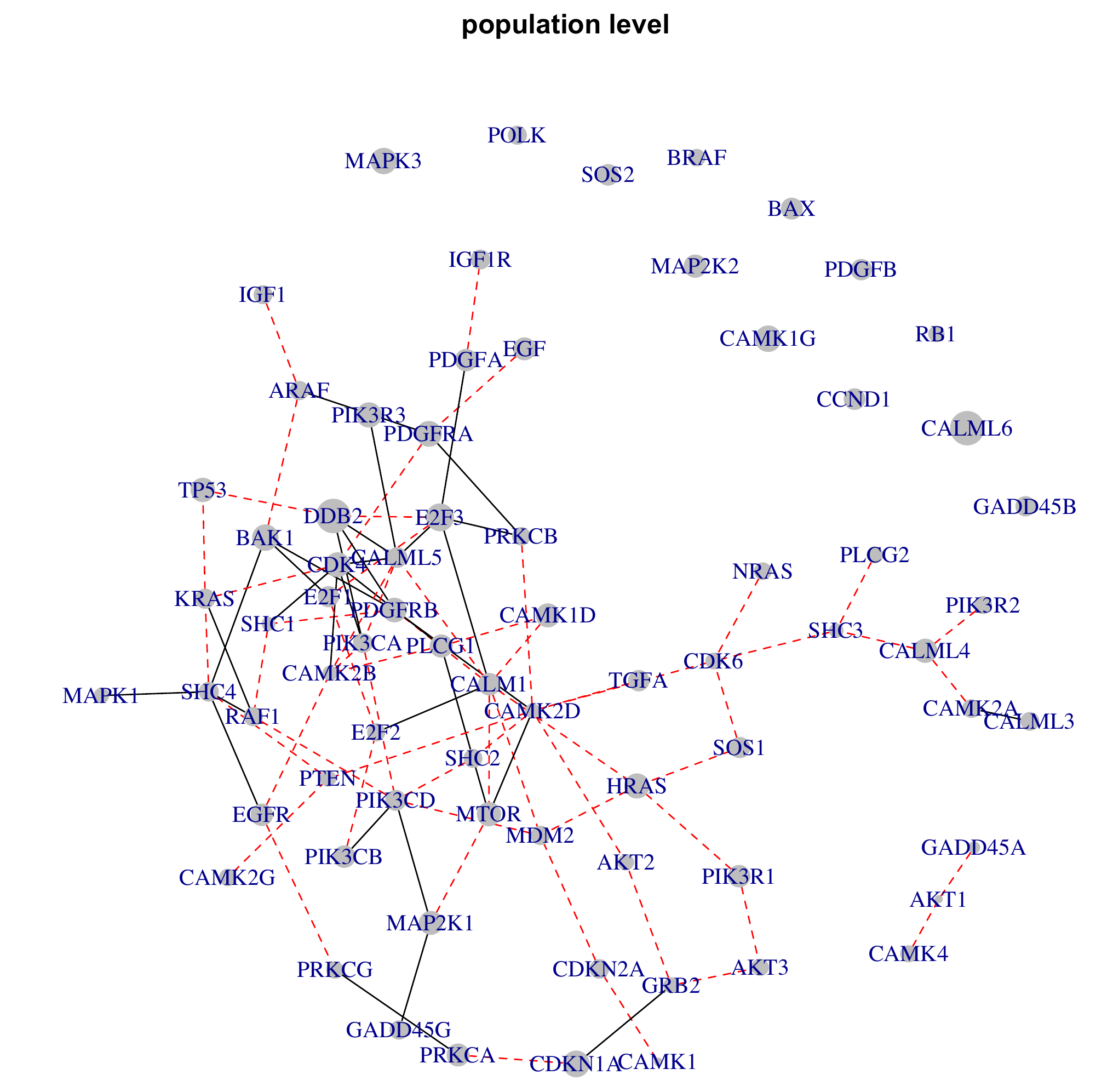}
\caption{The population-level  gene co-expression network, where the size of each node is proportional to the mean expression level, and edges with positive (negative) partial correlations are shown in red dashed (black solid) lines.}
\label{fig:pop}
\end{figure}
\begin{figure}[!t]
\centering
\includegraphics[trim=35mm 2cm 0 0, scale=0.35]{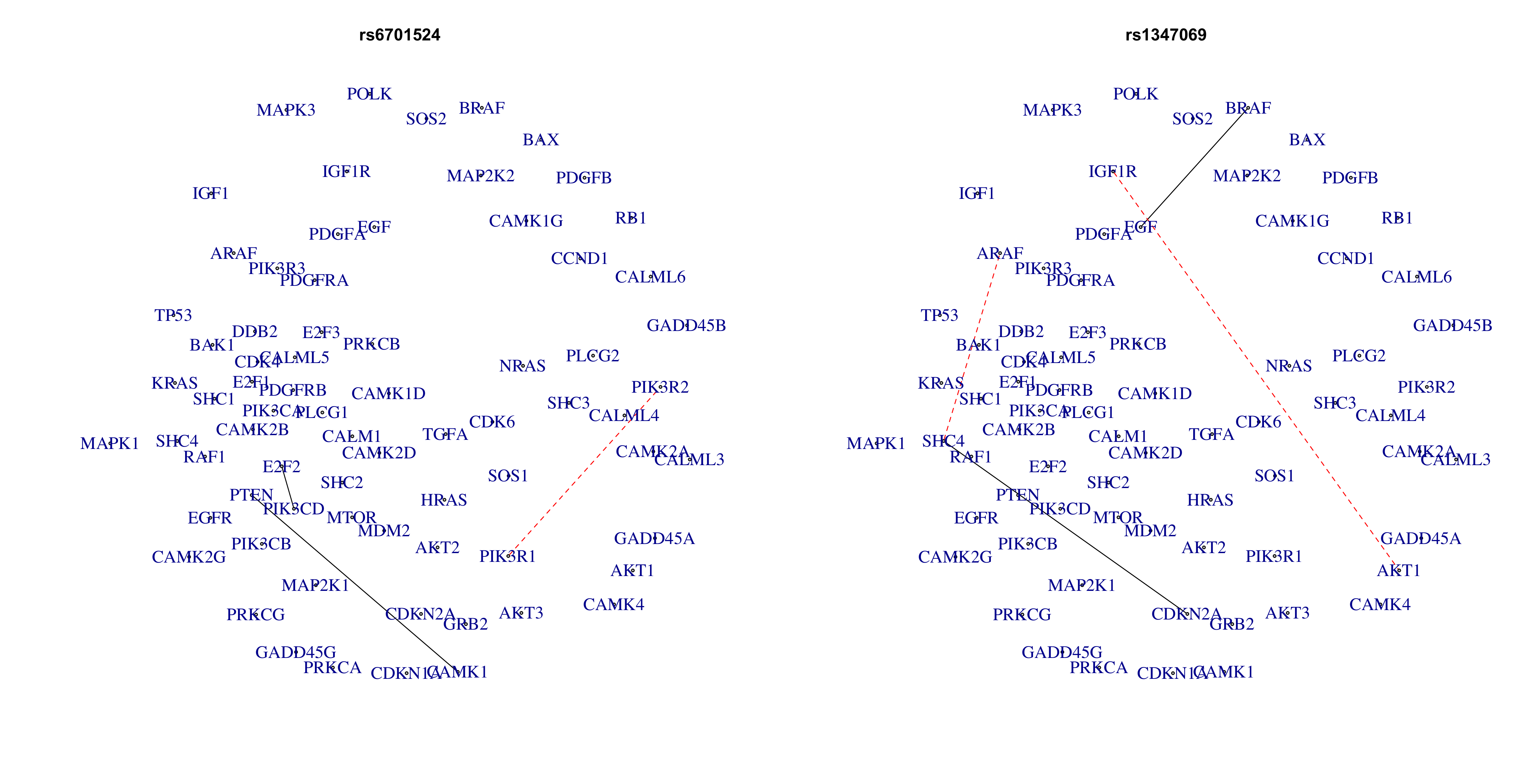}
\caption{Effects of each covariate (e.g., SNPs) on edges or partial correlations, with positive (negative) effects shown in red dashed (black solid) lines.}
\label{fig:cov}
\end{figure}

We have used the proposed method to construct the population level network as shown in Figure \ref{fig:pop}. Most of the connected genes in Figure \ref{fig:pop} are known to be oncogenes. For example, PIK3CA is a protein coding gene and is one of the most highly mutated oncogenes identified in human cancers \citep{samuels2004oncogenic}. The PIK3CA gene is a part of the PI3K/AKT/MTOR signaling pathway, which is one of the core pathways in human GBM and other types of cancer \citep{cancer2008comprehensive}. In Figure \ref{fig:pop}, we can identify several core pathways in human GBM including the PI3K/ AKT/MTOR, Ras-Raf-MEK-ERK and calcium signaling pathways.

We next examine the covariate effects on the network. Identified by our method are nine co-expression QTLs, namely, \texttt{rs6701524}, \texttt{rs10519201}, \texttt{rs1347069}, \texttt{rs9303511}, \texttt{rs503314}, \texttt{rs7286558}, \texttt{rs759950}, \texttt{rs306098}, \texttt{rs25919}. 
The network effects of \texttt{rs6701524} are shown in Figure \ref{fig:cov} (the left panel). 
This SNP, residing in MTOR, is found to affect co-expressions of genes in the PI3K/ AKT/MTOR pathway. This is an interesting finding as PI3K/MTOR is a key pathway in GBM development and progression, and inhibition of PI3K/MTOR signaling was found effective in increasing survival with GBM tumor \citep{batsios2019pi3k}.
This co-expression QTL can potentially play an important role in activating the PI3K/MTOR pathway. 
Shown in Figure \ref{fig:cov} (the right panel) are the network effects of \texttt{rs1347069}, a variant of MAP2K1. The figure indicates that this SNP mostly affects the co-expressions of genes in the Ras-Raf-MEK-ERK pathway. 

Moreover, we have found some novel co-expression QTLs. For example,
\texttt{rs10519201} regulates the co-expressions of PDGFB and GADD45A; 
\texttt{rs9303511} is associated with the co-expressions of GADD45A and CALML4; 
\texttt{rs503314} influences the co-expressions of CCND1 and PLCG2, CALML3; 
\texttt{rs7286558} may modify the co-expressions of EGFR and PRKCA; 
\texttt{rs759950} modulates the co-expressions of MAP2K1 and HRAS; 
\texttt{rs306098} influences the co-expressions of PIK3CD and CAMK2D; 
\texttt{rs25919}  may alter the co-expressions of CAMK2A and GADD45G. 
Co-expression QTL identification has recently sparked much interest, and these findings warrant  in-depth investigations.

\section{Conclusion}
We have proposed a computationally efficient multi-task learning estimator for Gaussian graphical regression models, and proved that the error rate of our  estimator improves upon that of the separate node-wise lasso estimator by a factor of $p$ (i.e., the dimension of a graph). The remarkable improvement is possible because multi-task learning borrows information across tasks.  Moreover, unlike the usual multi-task problems,  our tasks are entangled in a complicated correlation structure, defying existing  theoretical results; 
we  address it by establishing a new tail probability bound for correlated heavy-tailed variables with an arbitrary correlation structure. Simulations and a gene network application have suggested feasibility and utility of the proposal in high dimensional settings, where both the response variables and covariates exceed the sample size. Future work includes investigations of the optimal error rates and  the conditions needed to ensure variable selection consistency as well as extensions to accommodate more general types of responses, such as binary data.

\end{document}